\documentclass[a4paper, 10pt, twocolumn]{article} 
\usepackage[british]{babel}
\usepackage{microtype}
\usepackage{amsmath,amsfonts,amsthm}
\usepackage[svgnames,table]{xcolor}
\definecolor{GammaDarkGreen}{rgb}{0.113,0.278,0.2745}
\definecolor{GammaCharcoal}{rgb}{0.0706, 0.0706, 0.0706}
\definecolor{GammaRed}{HTML}{B71325}
\definecolor{GammaGreen}{HTML}{004846}
\usepackage{booktabs}
\usepackage{lastpage}
\usepackage{graphicx}
\usepackage{enumitem}
\setlist{leftmargin=*,itemsep=0.4ex,parsep=0ex,topsep=0.8ex,partopsep=0ex}
\setlist[itemize]{label=\color{GammaRed}$\triangleright$}
\usepackage{sectsty}
\chapterfont{\normalfont\Large\color{GammaRed}}
\sectionfont{\normalfont\Large\color{GammaRed}}
\subsectionfont{\normalfont\large\color{GammaCharcoal}}
\subsubsectionfont{\normalfont\normalsize\color{GammaGreen}}

\usepackage{subfig}

\usepackage{caption}
\DeclareCaptionLabelFormat{abbrevfig}{\color{GammaRed}Fig.\ #2}
\DeclareCaptionLabelFormat{abbrevtab}{\color{GammaRed}Tab.\ #2}
\usepackage[pdftex]{hyperref}
\pdftrailerid{}
\hypersetup{colorlinks=true,linkcolor=GammaDarkGreen,citecolor=GammaDarkGreen,filecolor=GammaDarkGreen,urlcolor=GammaDarkGreen}

\addto\extrasenglish{%
}

\usepackage{geometry}
\usepackage[utf8]{inputenc}
\usepackage[T1]{fontenc}
\IfFileExists{nimbusmononarrow.sty}{\usepackage{nimbusmononarrow}}{}
\usepackage{tgheros}
\usepackage{sansmath}
\usepackage{fancyhdr}
\usepackage{titling}

\newcommand{\authorstyle}[1]{{\large\color{GammaCharcoal}#1}}
\newcommand{\institution}[1]{{\footnotesize\itshape\color{GammaDarkGreen}#1\/}}

\pretitle{%
    \noindent
    \begin{minipage}[t]{0.63\textwidth}\vspace{20pt}
        \LARGE\color{GammaRed}
}

\posttitle{%
    \end{minipage}%
    \hfill
    \begin{minipage}[t]{0.33\textwidth}\vspace{-50pt}
        \raggedleft
        \includegraphics[width=\linewidth]{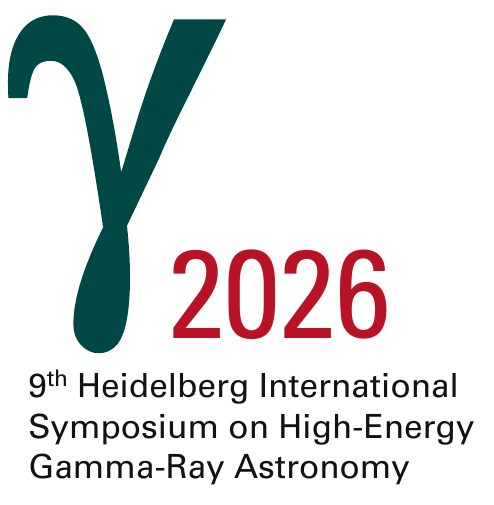}
    \end{minipage}%
}

\preauthor{\color{GammaDarkGreen}}
\postauthor{}
\predate{\begin{flushleft}\large\color{GammaRed}}
\postdate{\par\end{flushleft}}

\usepackage{lettrine}
\usepackage{xstring}
\newcommand{\initial}[1]{\lettrine[lines=2,findent=4pt,nindent=0pt]{\color{GammaRed}\emph{#1}}{}}
\newcommand{\lettrineabstract}[1]{\StrLeft{#1}{1}[\firstletter]\initial{\firstletter}\emph{{\StrGobbleLeft{#1}{1}}}}

\usepackage[backend=bibtex,style=numeric-comp,sorting=none,giveninits=true,citestyle=numeric]{biblatex}
\usepackage[autostyle=true]{csquotes}

\usepackage{lipsum}  

\title{Radio Galaxies as UHECR Sources in Light of Recent Microquasar Results}

\author{
	\authorstyle{A.~M.~Taylor\textcolor{GammaDarkGreen}{\textsuperscript{1}}, J.~H.~Matthews\textcolor{GammaDarkGreen}{\textsuperscript{2}}}\\
\newline
    {\footnotesize
        \textsuperscript{\color{GammaDarkGreen}1}\institution{Deutsches Elektronen-Synchrotron, Platanenallee 6, D-15738 Zeuthen, Germany}\\
\textsuperscript{\color{GammaDarkGreen}2}\institution{Department of Physics, Astrophysics, University of Oxford, Denys Wilkinson Building, Keble Road, Oxford, OX1 3RH, UK}
}
}

\newcommand{\website}{\href{https://plan.events.mpg.de/event/543/}{Proceedings of the 9th Heidelberg Gamma-Ray Symposium}}

\begin{document}
\sansmath\maketitle
\thispagestyle{fancy}

\lettrineabstract{The recent discovery that microquasars operate as efficient particle accelerators brings
back into focus the question as to whether their larger cousins, radio galaxies, also operate as
efficient accelerators. We here discuss what we presently know about particle acceleration in radio
galaxies, focusing particularly on the nearest such object, Centaurus~A.}

\section{Introduction}

\begin{figure}
    \centering
    \includegraphics[width=8cm]{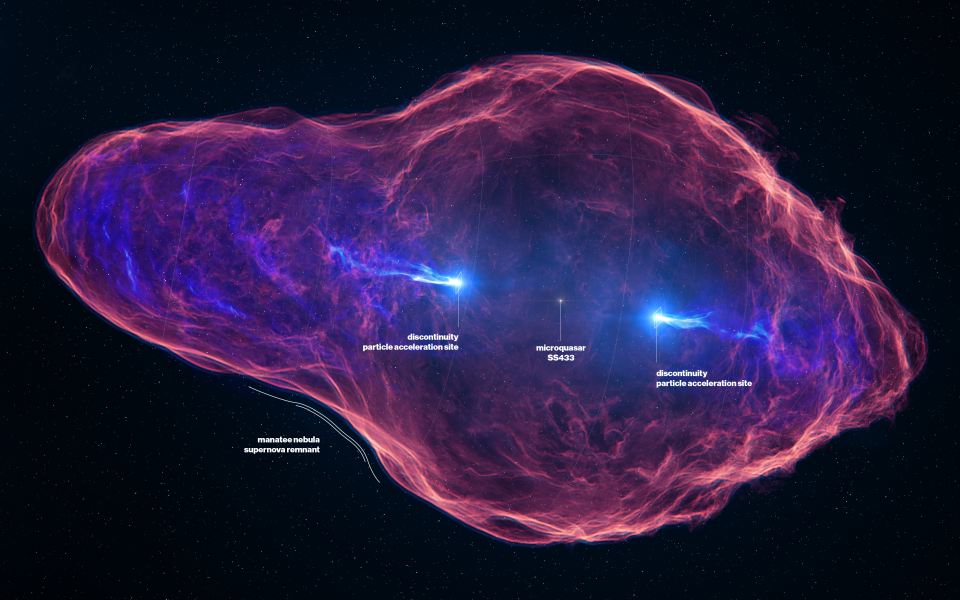}
    \caption{An image of the Galactic microquasar SS~433: Science Communication Lab for MPIK/H.E.S.S.}
    \label{fig:SS433}
\end{figure}

Recent developments from the study of black hole systems which produce powerful jets have provided fresh insights into
the ability of these systems to operate as particle accelerators.
Specifically, both X-ray and very high energy $\gamma$-ray observations of Galactic microquasars, stellar
mass black hole systems with a companion star and jetted outflow, have revealed that the jets from these objects
are capable of accelerating particles extremely efficiently \cite{2024Sci...383..402H,2025NSRev..12af496L}. 
A fiducial example object of this source class is SS~433 (see Fig.~\ref{fig:SS433}). 
The efficiency of particle acceleration in the jets, and the corresponding implications this has on the maximum 
particle energy achievable by these sources, naturally raise the question as to whether these sources can significantly
contribute to the observed Galactic cosmic ray flux at Earth \cite{2026ApJ..1006..211A}.

\begin{figure}
    \centering
    \includegraphics[width=8cm]{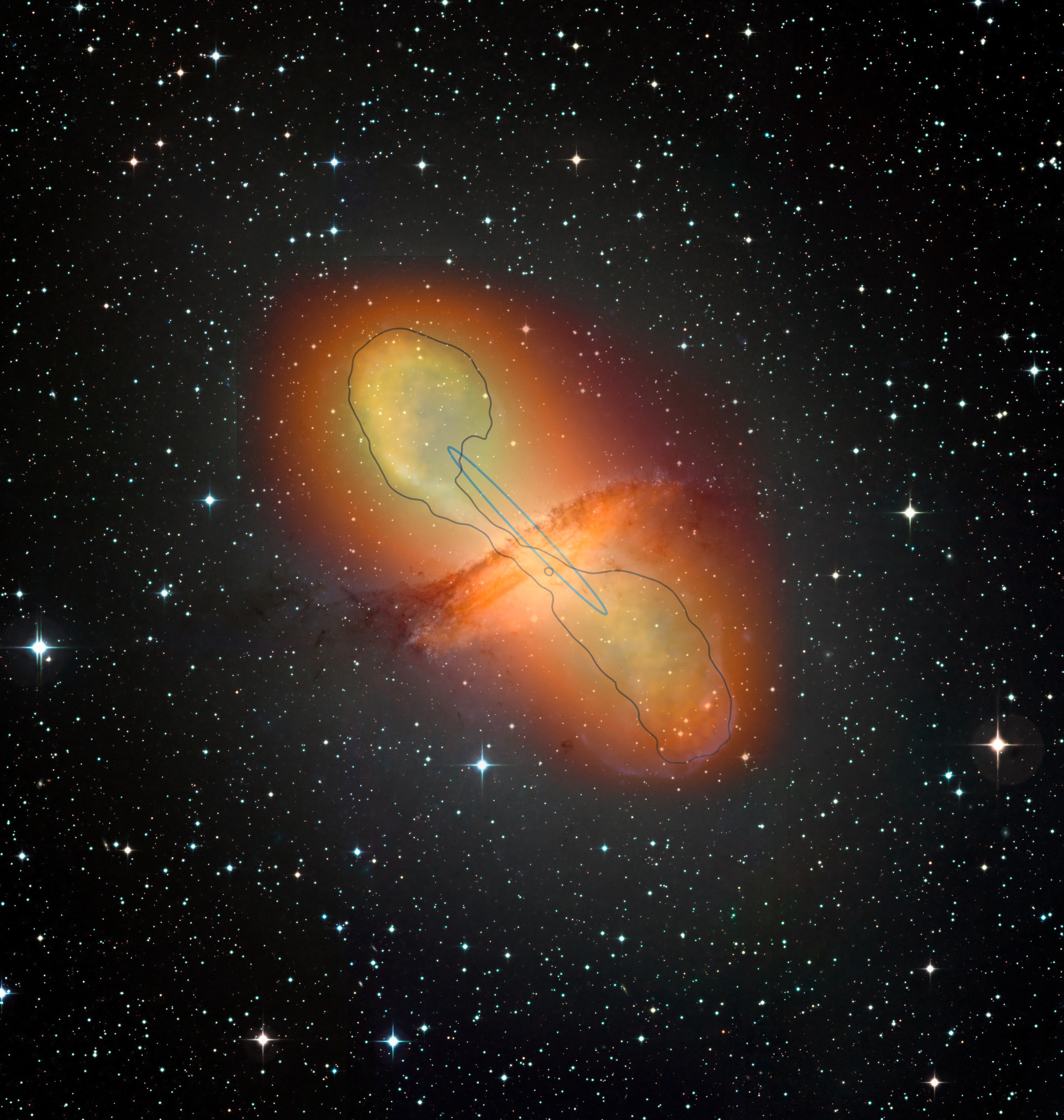}
    \caption{An image of the local radio galaxy Cen~A: Science Communication Lab for MPIK/H.E.S.S.}
    \label{fig:CenA}
\end{figure}

These revelations from Galactic jetted black hole systems have ramifications even beyond the role of microquasars 
as Galactic CR sources. They also bring back into focus the question whether supermassive 
black hole system with jets, more luminous extragalactic cousins of microquasars, may also be able to 
operate as efficient accelerators. A fiducial example object of this source class is the local radio galaxy, 
Centaurus~A (from here on referred to as Cen~A) (see Fig.\ref{fig:CenA}).

\section{Synchrotron Emission Probe of UHECR Source Candidates}
\label{Sync_Probe}

The particle acceleration timescale for particles accelerated by an astrophysical source can generally be expressed as,
\begin{eqnarray}
t_{\rm acc}=\eta \frac{R_{\rm Lar}}{\beta^{2}c},
\end{eqnarray}
where $R_{\rm Lar}$ is the particle Larmor radius within the source's magnetic field, $\beta c$ is the velocity of the 
scatterer giving rise to the acceleration of the particles (comparable in magnitude to the velocity of the flow), and 
the parameter $\eta$ describes how close to the Bohm 
level the particle pitch angle scattering takes place in the accelerator. The corresponding escape time for hadrons 
from the acceleration site, assuming advective escape is the dominant process,
\begin{eqnarray}
t_{\rm esc}=\frac{R}{\beta c},
\end{eqnarray}
where $R$ is the associated source size. By equating these two timescales, the maximum energy 
hadron achievable within a source is found (the Hillas limit \cite{1984ARA&A..22..425H}),
\begin{eqnarray}
E_{\rm max}=\beta \eta^{-1} Ze B R.
\end{eqnarray}

This limit in maximum energy achievable by a source can be more directly connected to candidate 
source observables by assuming equipartition of the source magnetic and kinetic energy luminosities, 
leading to a relation between the 
source kinetic energy luminosity, $L_{\rm KE}$, and the maximum energy $E_{p, {\rm max}}$ \cite{1995ApJ...454...60N},
\begin{eqnarray}
L_{\rm KE}>3\times 10^{43}\frac{\eta^{2}}{\beta}\left(\frac{E_{p, {\rm max}}}{10~EeV}\right)^{2}~{\rm erg~s}^{-1}.
\label{UHECR_Source_Luminosity}
\end{eqnarray}
This result highlights that candidate UHECR sources require not only large kinetic energy luminosities and 
large associated outflow velocities (i.e. $\beta\sim 1$), but also Bohm level scattering times (i.e. $\eta\sim 1$).

Further insight into such candidate UHECR sources can be provided from the consideration of electron
acceleration by the same object. Unlike the highest energy hadrons which may leave the acceleration
region via advective escape from the accelerator, the maximum energy for electrons is instead set by their energy 
loss timescale within the accelerator. The energy loss time for electrons within a source, assuming it is dominated by 
synchrotron losses, can be expressed as,
\begin{eqnarray}
t_{\rm loss}=\frac{9}{8\pi\alpha}\frac{m_{e}}{E_{\gamma}^{\rm sync}}t_{\rm Lar}
\end{eqnarray}
where $\alpha$ is the fine structure constant, $E_{\gamma}^{\rm sync}$ is the synchrotron emission 
energy scale, and $t_{\rm Lar}$ is the particle Larmor period timescale. Equating this timescale with the 
acceleration timescale, the maximum energy synchrotron emission that can be produced by a source
is,
\begin{eqnarray}
E_{\gamma, {\rm max}}^{\rm sync}=\frac{9}{4}\frac{\beta^{2}}{\eta}\frac{m_{e}c^{2}}{\alpha}
\end{eqnarray}
where $m_{e}c^{2}$ is the electron rest mass energy. One can therefore probe particle acceleration efficiency
within a source through the determination of the maximum energy synchrotron emission achieved by a source 
before the onset of the cutoff in this component.

\subsection{Particle Acceleration Efficiency in the jets of SS~433 and the Inner Jets of Cen~A}


\subsubsection{SS~433}
A comparatively local and well-resolved Galactic black hole jet system
is the microquasar SS~433. This binary system is composed of a stellar-mass black hole, 
a stellar companion, and fast collimated outflow. Although the value for 
the composition ratio (i.e. the lepton-to-hadron ratio) of SS~433's jet remains unclear, 
evidence for a hadronic particle component in the jet has been found \cite{2002Sci...297.1673M}.
The jetted outflow from the black hole is observed to give
rise to bright non-thermal emission at various regions along the jet on scales of 10s of pc
from the black hole (see Fig.~\ref{fig:SS433}).

In order to probe the acceleration efficiency of the particles giving rise to this non-thermal emission,
consideration must be made of the acceleration and cooling rates of electrons in the source.
Of the two cooling channels, synchrotron cooling was shown to significantly dominate over
inverse Compton cooling \cite{2020ApJ...889..146S}. As discussed in Sec.~\ref{Sync_Probe}, 
analysis of the shape of the cutoff in the synchrotron spectrum, produced by the 
highest energy electrons, directly constrains a combination of both the source acceleration 
efficiency parameter, $\eta$, and the velocity of the scatterers, $\beta$, namely $\eta \beta^{-2}$.

An investigation of the shape of the synchrotron spectrum in \cite{2020ApJ...889..146S} 
found the combined constraint, $\eta \beta^{-2}<10^{2}$. Given that the outflow velocity of the
jet is around $0.26~c$ \cite{1989ApJ...347..448M}, this constraint suggests that $\eta<10$ for this source, 
namely it is operating as an extremely efficient particle accelerator.


\subsubsection{Cen~A Inner Jets}
Likewise, Cen~A is a local and spatially well-resolved radio galaxy.
Interestingly, the inner jets of Cen~A on $\sim {\rm kpc}$ distance scales
are estimated to harbour similar-strength magnetic fields ($\sim 20~\mu$G) to that possessed by SS~433 on 
$\sim 10~{\rm pc}$ scales \cite{2020Natur.582..356H,2020ApJ...889..146S}.
In contrast to this coincidental similarity in the magnetic field levels of these systems, 
the level of the radiation field within the emission zone for 
Cen~A's inner jets sits close to the equipartition level with the magnetic field. 
This ratio of electromagnetic fields in the source is unlike the emission zone of SS~433's jets, 
whose magnetic fields dominate significantly over the radiation field.
Consequently, the levels of synchrotron and inverse
Compton emission produced by Cen~A's inner jets are of comparable magnitude, limiting the ability to probe the cutoff
of the synchrotron emission. Subsequently, a much weaker constraint of $\eta \beta^{-2}<10^{4}$, 
compared to that obtained for SS~433, can presently be placed on the inner jets of Cen~A. Given 
the relativistic outflow velocity of these jets, this constrains the acceleration efficiency 
to $\eta< 10^{4}$.

\begin{figure*}
    \centering
    \includegraphics[width=15cm]{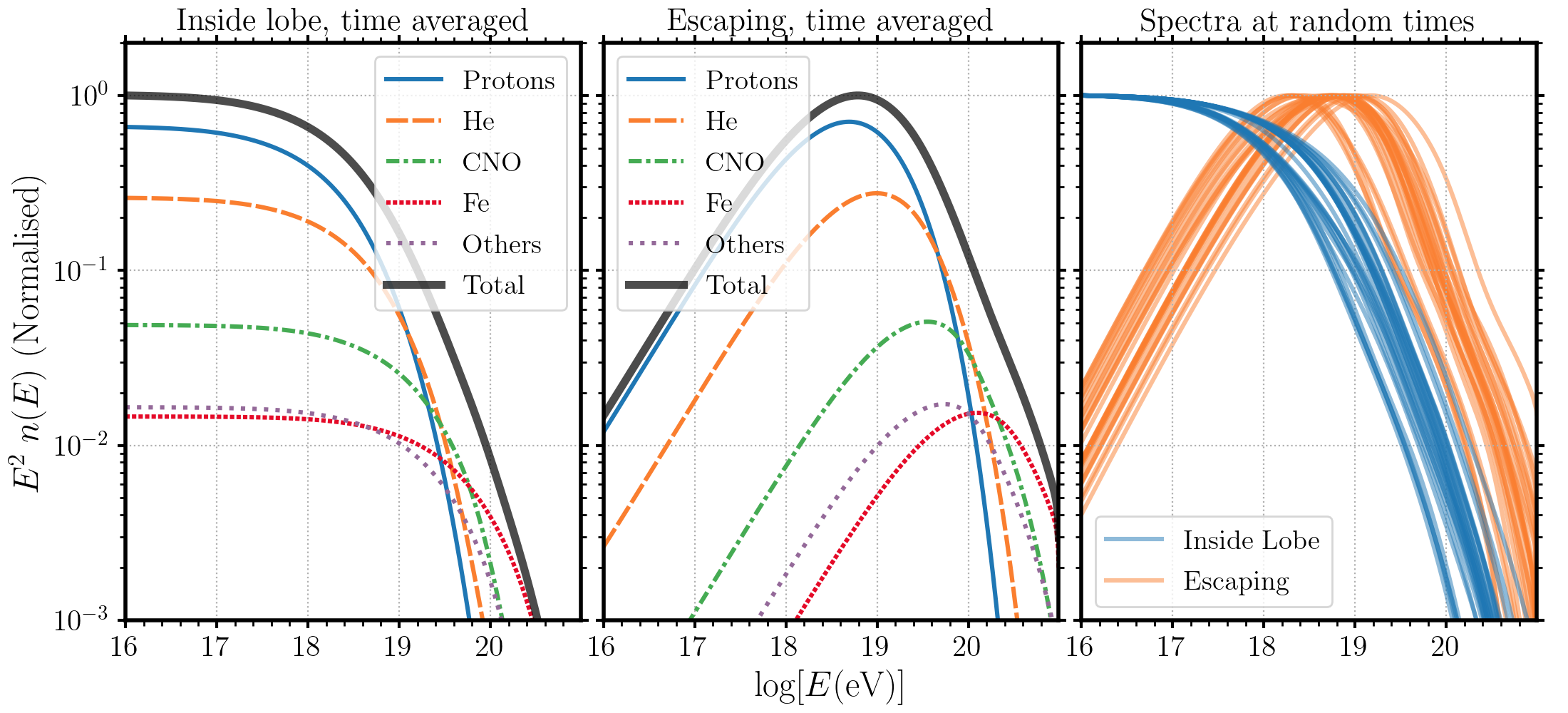}
    \caption{Left-panel: the steady state hadronic particle spectrum found within the cocoon region. Middle-panel: the steady-state hadronic particle spectrum found for the escaping particles from the cocoon. Right-panel: an ensemble of random snapshots of the hadronic particle spectrum both within the cocoon and outside the cocoon region. This figure is from \cite{2021MNRAS.503.5948M}.}
    \label{fig:hadron_spec}
\end{figure*}

\subsection{The Synchrotron Cutoff from the Outer Jets of Cen~A}
\label{Synch_Cutoff_CenA}

Out along Cen~A's jets beyond the inner jet region, its outer jets 
are observed to extend out to 100s of kpcs in distance. 
The fact that both the inner and outer jet structures are both distinctly observed, by itself,
provides direct evidence of varied emission of the jets of Cen~A over the timescales needed to create these 
structures. The dynamical timescales associdated with the production of these inner and outer jet 
structures is estimated to be Myr and 100-1000 Myr, respectively \cite{2026MNRAS.545f2160D}. Their
existence, therfore, indicates that the luminosity of Cen~A's jets has varied on a timescale
between these two scales.


The synchrotron emission from both the northern and southern outer jet regions reveal spatially
dependent spectral evolution \cite{2009MNRAS.393.1041H,2016A&A...595A..29S}. In particular,
evidence for either a cooling break (suggested to be present in the northern lobe emission spectrum by \cite{2009MNRAS.393.1041H}) 
or cooling cutoff (suggested to be present in the southern lobe emission spectrum by \cite{2009MNRAS.393.1041H}) 
is found for these two lobes from their spatially dependent spectral measurements. 
With radiative cooling through interactions with the CMB dominating
over synchrotron cooling out at these large distances, the location of the cooling break can be 
converted into a cooling time. The cooling time inferred, 30~Myr \cite{2009MNRAS.393.1041H}, is 
considerably shorter than the estimated dynamical time estimated for creation of the outer jet
structures.

Such an order of magnitude overestimate of the dynamical age compared to the spectral age of radio galaxies is 
not unique to Cen~A, but is rather a trend generally found for these objects through the application of these two age estimates
\cite{1996ASPC..100..281E}. The two main options for the resolution of this mismatch in ages, dependent on the 
method employed for the age determination \cite{1996ASPC..100..281E,2026arXiv260628755H}, are:
\begin{enumerate}
\item source variability 
\item multiple acceleration sites along the jet
\end{enumerate}

\section{Episodic AGN Simulations \& Proxy Electrons}


In order to investigate the effects of AGN jet luminosity on their non-thermal particle signatures,
a simplified simulation setup was adopted \cite{2021MNRAS.503.5948M}. In these simulations:
\begin{enumerate}
\item particle acceleration was assumed to only take place at the jet termination shock with $\eta=1$
\item the accelerated particles subsequently diffuse isotropically throughout the inflated cocoon volume
\item diffusion throughout the cocoon is assumed to operate at the Bohm level.
\end{enumerate}

In these simulations, the jet luminosity was varied in a probabilistic manner, following an assumed 
log-normal luminosity probability density function,
with a pink noise power spectrum density. Hadrons accelerated by the jets, upon diffusing through the
cocoon volume, subsequently escaped through the cocoon surface. Following simulation runs under these assumptions, snapshots of the 
quasi-steady state hadronic particle spectrum were obtained. As seen from Fig.~\ref{fig:hadron_spec}, during epochs with
sustained jet power at or beyond $10^{44}~{\rm erg~s}^{-1}$, acceleration up to 10~EeV is found to be possible.

\begin{figure*}
    \centering
    \includegraphics[width=10cm]{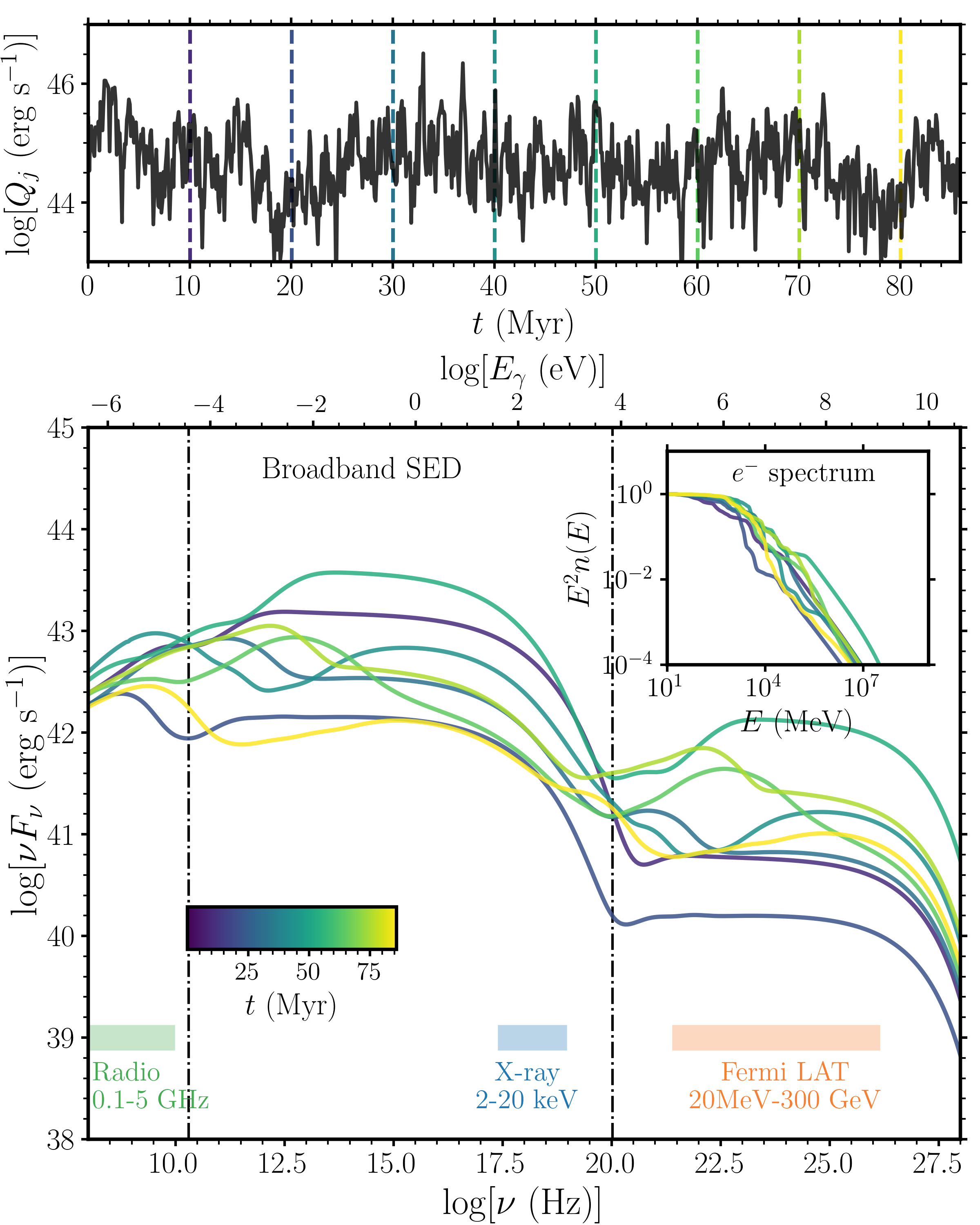}
    \caption{Top panel: the variation of jet luminosity with time adopted in the simulation, with 10~Myr snapshot regions indicated. Bottom panel inset: quasi-steady state electron spectrum. Bottom panel: quasi-steady state emission from electrons. The bashed vertical lines indicate the emission from ``proxy electrons'' whose cooling times match the escape time of UHECR hadrons. This figure is from \cite{2021MNRAS.503.5948M}.}
    \label{fig:electron_spec}
\end{figure*}

For electrons also accelerated by the jets, particle energy cooling rather than escape dominates the loss term in the 
momentum transport equation. A set of snapshot results from the simulation, obtained 
for the electron quasi-steady state spectrum, is shown in Fig.~\ref{fig:electron_spec}. Also indicated in the figure (see
dashed vertical lines) are the emission energy regions from ``proxy electrons'', whose energy loss cooling times match the
UHECR hadron escape times. This result highlights the importance of studying AGN jets in these particular frequency bands
(namely around 10~GHz and 10~keV) in order to probe the source activity on timescales of relevance for UHECR acceleration 
and escape.

\section{Multiple Acceleration Sites}

On top of evidence for temporal variability, observational evidence for multiple acceleration sites along the jet
has also been found. Specifically, Chandra X-ray observations of Cen~A's jets on kpc scales
\cite{2007ApJ...670L..81H} have revealed both distinct knot emission regions along the jet as well as more
general diffuse emission.
In the VHE $\gamma$-ray band, HESS observations have also resolved non-point source emission, 
revealing that this emission is instead extended along the direction of the jet \cite{2020Natur.582..356H}.

However, whether the extended VHE emission relates to knots or a diffuse origin remains unclear.
Both options have interesting implications for particle acceleration.
Assuming the acceleration site relates to knot regions, \cite{2025MNRAS.539.3697D} estimate what
the potential maximum energy is for protons in the source. Under favourable assumptions, these 
estimations found that energies up to $E_{p,{\rm max}}=4\times 10^{17}$~eV were potentially achievable. 
However, such an estimate was noted to be extremely optimistic, and factoring in levels of realism into 
the estimate brought this maximum energy estimate down to closer to PeV energies.

Alternatively, if the acceleration site is extended and distributed over a broad region, this would 
indicate that the acceleration 
relates to processes which operate continuously along the jet such as the shear acceleration
mechanism \cite{2019Galax...7...78R}. In this process, non-thermal particles moving in the magnetic
fields of the AGN jet, probe the velocity gradient across the jet on length scales of their Larmor radius.
As a consequence the acceleration time of the particles at low energies is naively expected, 
at least within the framework of quasi-linear-theory, to be longer than at high energies, primarily as a 
result of the differing energy boost received in the scattering process for different energy particles.

However, the actual scattering time and energy boost received by particles in the shear acceleration process is 
sensitive to the magnetic field structure present. As highlighted in \cite{2022icrc.confE.491O},
the shearing process can modify the magnetic field structure within the shear layer. Consequently, this
modification alters the particle transport rates across the shear layer, leaing to a significant departure 
for the acceleration time of particles compared to the expectations from quasi-linear-theory,
particularly for the lowest energy particles.
For the highest energy particles accelerated by the jet, with Larmor radii comparable 
to the to the jet radius, more modest modifications to the expected acceleration times are
expected. The shear acceleration process, therefore, offers a promising mechanism for efficient
particles acceleration in jets up to the ultra high energy scale.


\section{UHECR as Messengers About Their Sources}

Any candidate UHECR source class must be able to account for the ensemble of observational UHECR results.
Specifically, models adopting a given source class must be able to satisfactory simultaneously describe the 
UHECR energy spectrum, composition, and energy dependent skymaps.

Focusing initially on only combined fits to the energy spectrum and composition data, under the assumption that all sources have
the same spectral slope, cutoff rigidity, and composition, the landscape of source parameters has previously
been explored \cite{2015PhRvD..92f3011T,2017JCAP...04..038A}. These explorations found that the source spectral index was 
sensitive to the assumed spatial evolution of the sources adopted. A source distribution density which increased with redshift
favoured hard source spectra, whereas distribution densities which decreased with redshift favoured soft source
spectra. As was noted in \cite{2017JCAP...04..038A}, for the case in which the sources reside only locally ($z<0.02$,
ie. the case of a dominant local source \cite{2016PhRvD..94d3008L}), a particular soft source spectra was shown to be required.

These findings are particularly striking when juxtaposed with more recent results which have highlighted the 
sensitivity that these fit solutions have on the source spectra assumptions \cite{2023PhRvD.107j3045E}.
Specifically, it was demonstrated that the hard spectra results are only viable provided that all sources have close to 
the same rigidity cutoff.
It therefore seems warranted to consider more seriously the possibility of the dominance of a local source scenario,
and in particular whether such a scenario is compatible with the anisotropy structure revealed in \cite{2024icrc.confE.521T}. 

\begin{figure*}
    \centering
    \includegraphics[width=10cm]{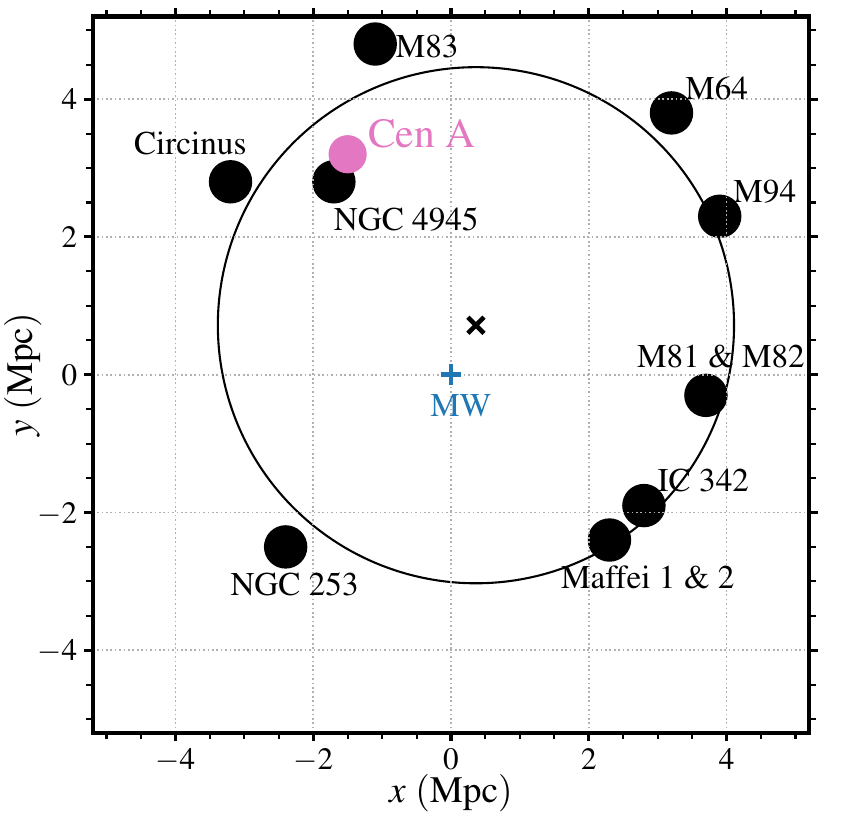}
    \caption{The location of local extragalactic galaxies (the ``Council of Giants''), with positions provided within the Local Sheet coordinate system \cite{2014MNRAS.440..405M}. This figure is from \cite{2023MNRAS.524..631T}.}
    \label{fig:CoG}
\end{figure*}

\begin{figure*}
    \centering
    \includegraphics[width=15cm]{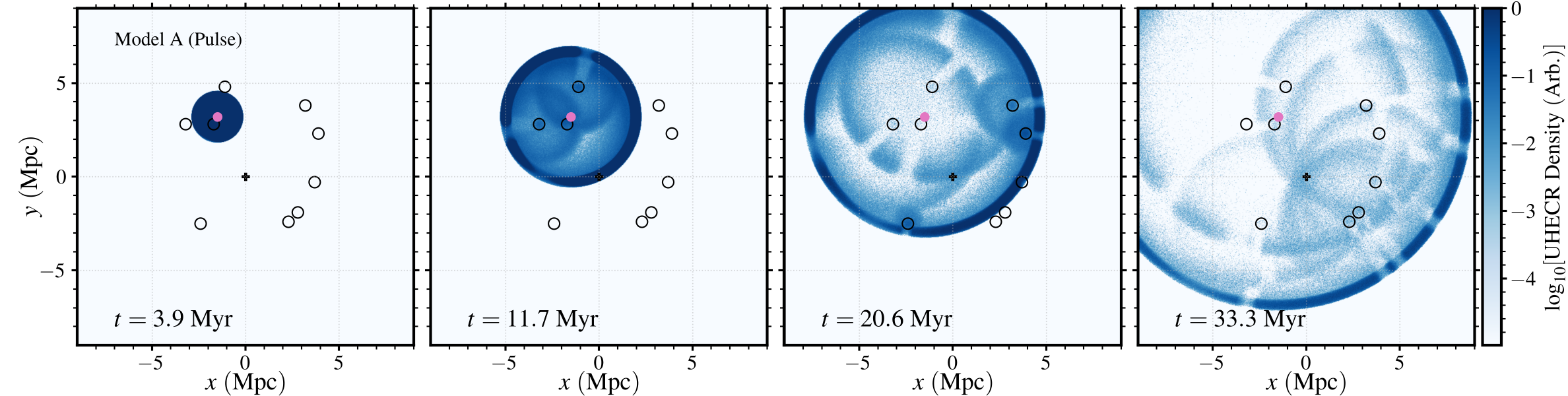}
    \caption{A set of snapshots from the simuation, taken at key timesteps when different sets of echo waves arrive to the Milky Way. This figure is from \cite{2023MNRAS.524..631T}.}
    \label{fig:Arrival_Times}
\end{figure*}

\begin{figure*}
    \centering
    \includegraphics[width=10cm]{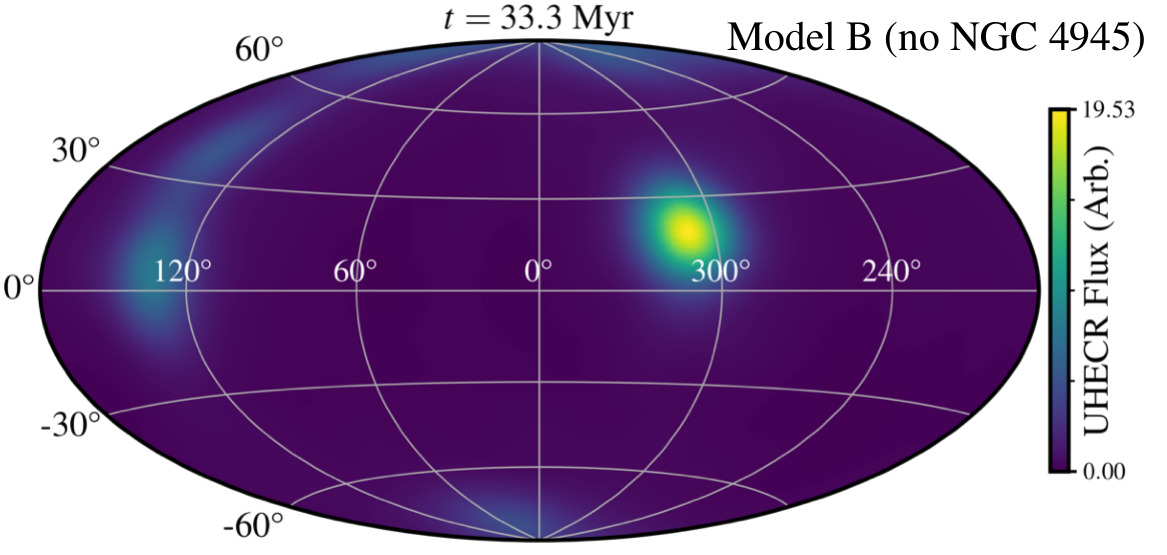}
    \caption{The arrival skymap of cosmic rays obtained from the similation at 33.3~Myrs after Cen~A's outburst event. This figure is from \cite{2023MNRAS.524..631T}.}
    \label{fig:Arrival_Skymap}
\end{figure*}

A level of correlation of the arriving distribution of UHECR with the supergalactic plane has previously been noted \cite{2025ApJ...984..123A}.
This plane closely aligns to the plane of the Local Sheet, defined by the plane of galaxies in the local extragalactic neighbourhood, commonly referred to as the Council of Giants (see Fig.~\ref{fig:CoG}). Within this set of objects, Cen~A is the only member whose energetics are close to satisfying the luminosity requirements for a candidate UHECR source (see Eqn.~\ref{UHECR_Source_Luminosity}).

Within the framework of the local source scenario, it therefore appears reasonable to ask whether UHECR may originate from the 
Council of Giants. This question was considered in \cite{2022MNRAS.511..448B,2023MNRAS.524..631T}, where it was noted that only Cen~A
appears a viable UHECR source within this group. However, a significant imprint of the Council of Giants structure was shown to be 
expected in a scenario in which Cen~A underwent a outburst episode around 30~Myrs ago (as was already motivated to be the 
case, see Sec.~\ref{Synch_Cutoff_CenA}), provided that the extended haloes around these galaxies are sufficiently magnetised.

\section{Conclusion}
Recent observations of Galactic solar mass black hole systems with jetted outflows, namely microquasars, indicate 
that these systems operate as extreme accelerators, accelerating particles beyond the PeV energy scale. This result 
motivates a return to the consideration of extragalactic supermassive black hole systems with jetted outflows,
namely radiogalaxies, as multi-EeV energy particle accelerators. 

The synchrotron emission from the sources, in both cases, is used as a probe of the source acceleration
efficiency. For microquasars, acceleration at the Bohm limit ($\eta\approx 1$) is found to be achieved.
For radiogalaxies, only a weaker constraint of $\eta<10^{4}$ is found.

The consideration of the shape of the synchrotron spectrum for radiogalaxies, however, provides evidence
that these source exhibit considerable variability, with the local radiogalaxy, Cen~A, demonstrating that 
the object has been much more active in the recent past ($>30$~Myr).

The possibility that a local source, which underwent an active episode in the recent past, is able to
explain a multitude of facets of UHECR data. Specifally, it can account for the spectrum and composition
observed, requiring a considerably softer source spectrum than that required for more distant source 
scenarios. Such a scenario can also account for the imprint of local extragalactic structure observed
in the arriving UHECR skymap.

\printbibliography[title={References}]

\end{document}